\documentclass[conference]{IEEEtran}

\usepackage{bbm}

\ifCLASSINFOpdf
  \usepackage[pdftex]{graphicx}
\else
  \usepackage[dvips]{graphicx}
\fi
\usepackage[cmex10]{amsmath}
\usepackage{amssymb}
\usepackage[]{algorithm2e}
\usepackage{amsfonts}
\usepackage[]{algorithm2e}
\usepackage{cases}
\usepackage{array}
\usepackage[caption=false,font=footnotesize]{subfig}
\begin{document}
%
\title{Run Time Approximation of Non-blocking Service Rates for Streaming Systems}

\author{\IEEEauthorblockN{Jonathan C. Beard and Roger D. Chamberlain}
\IEEEauthorblockA{Dept. of Computer Science and Engineering\\
Washington University in St. Louis\\
One Brookings Drive\\
St. Louis, Missouri 63130\\
Email: \{jbeard,roger\}@wustl.edu}
}


%


\maketitle

\begin{abstract}
Stream processing is a compute paradigm that promises safe and
efficient parallelism.  Modern big-data problems are often well
suited for stream processing's throughput-oriented nature.
Realization of efficient stream processing requires monitoring
and optimization of multiple communications links.  Most techniques
to optimize these links use queueing network models or network flow
models, which
require some idea of the actual execution rate of each 
independent compute kernel within the system.  What we want to know is 
how fast can each kernel process data independent of other
communicating kernels. This is known as the ``service rate'' 
of the kernel within the queueing literature.  Current approaches to divining
service rates are static.  Modern workloads, however, are often 
dynamic.  Shared cloud systems also present applications with 
highly dynamic execution environments (multiple users, hardware
migration, etc.).  It is therefore desirable
to continuously re-tune an application during run time 
(online) in response to changing conditions.
Our approach enables online service rate monitoring
under most conditions, obviating the need for reliance on steady state predictions for 
what are probably non-steady state phenomena.  First, some of the 
difficulties
associated with online service rate determination are examined. Second,
the algorithm to approximate the online non-blocking service rate
is described.  Lastly, the algorithm is implemented within the 
open source RaftLib framework for validation using a simple 
microbenchmark as well as two full streaming applications.

\end{abstract}


%
\IEEEpeerreviewmaketitle

\section{Introduction}\label{sec:intro}

Stream processing (or data-flow programming) is a compute paradigm
that enables parallel execution of sequentially constructed 
kernels.  This is accomplished by managing the flows of data
from one sequentially programmed  kernel to the next.  Flows 
of data within the stream processing community are known 
as ``streams.'' Queueing behavior naturally arises between 
two kernels independently executing. Selecting the correct queue 
capacity (buffer size) is
one parameter (of many) that can be
critical to the overall performance of the 
streaming system.  Doing so, however,
often requires information, such 
as the service rate of each kernel,
not typically available at run time (online).  
Complicating matters further for online optimization, many analytic 
methods used to solve for optimal buffer size require an 
understanding of the underlying service process distribution,
not just it's mean.  
Both service rate and process distribution
can be extremely difficult to determine
online without effecting the behavior of the application (i.e., degrading
application performance).  This paper proposes and demonstrates 
a heuristic that enables online 
service rate approximation of each compute kernel within
a streaming system.  It is also 
shown that this method imposes minimal impact on the monitored
application.  

\begin{figure}[ht]
\centering
\includegraphics[width=2.82in,height=1.17in]{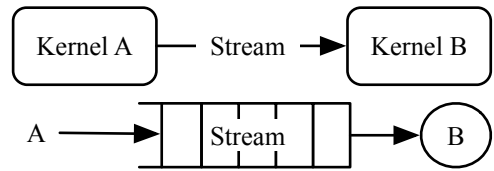}
\caption{The simple streaming application at top has two
compute Kernels A \& B with a single stream connecting them.  The
corresponding queueing network is a single server B with a single queue
fed by the arrival process generated by Kernel A.}
\label{fig:microbenchmark}
\end{figure}

An example of a simple streaming application is shown in 
Figure~\ref{fig:microbenchmark}.  The kernel labeled as ``A''
produces output which is ``streamed'' to kernel ``B'' over
the communications link labeled ``Stream.''  These communications links
are directed (one way).  Strict stream
processing semantics dictate that all of the state necessary for each 
kernel to operate is compartmentalized within that kernel,
the only communication allowed utilizes the stream.  State 
compartmentalization and subsequent one-way transmittal of 
state via streaming comes with increased communications between
kernels.  Increased communication comes with multiple costs
depending on the application: increased latency, decreased throughput,
higher energy usage.  No matter what the cost 
function is, minimizing its result often involves optimizing
the streams (queues and subsequent buffers)  connecting individual 
compute kernels.  The buffers forming the streams of the 
application can be viewed as a queueing 
network~\cite{bc13b,lavenberg1989perspective}.
It is this network that
we want to optimize while the application is executing.

Optimizing the queueing network that models a streaming application
can be performed using analytic techniques.  Ubiquitous to many of these
models is the non-blocking service rate of each compute kernel.  Classic
approaches assume a stationary distribution.  This carries
the assumption that both the workload presented to the
compute kernel and its environment are stable over time.  One only has to 
look at the variety of data presented to any common application
to realize that the assumption of a persistent homogeneous workload
is naive.  With the popularity of cloud computing we also have
to assume that the environment an application is executing in
can change at a moments notice, therefore we must build applications
that can be resilient to perturbations in their execution
environment.  We focus
on low overhead instrumentation that will enable more resilient 
stream processing applications by informing the runtime when conditions 
change.

\begin{figure}[ht]
\centering
\includegraphics[height=2.5in,width=3in]{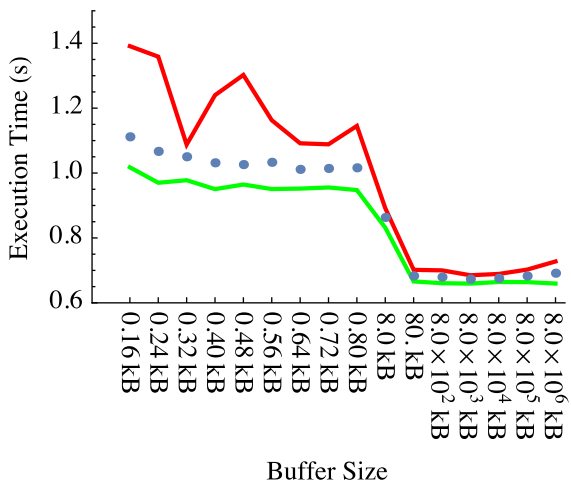}
\caption{Incorrect buffer sizes can have a deleterious effect
on the overall performance of a streaming system.  This chart
depicts empirical data collected from a matrix multiply application
executing on eight threads.  The points represent mean execution
time, and the lower and upper
solid lines are the $5^{th}$ and $95^{th}$ percentiles, respectively.
The times are for the overall execution
of the algorithm only, no allocation or deallocation time included.  The net effect
of increased buffer sizes initially improves performance but then 
slowly degrades as the buffer size is increased.}
\label{fig:toobigalloc}
\end{figure}

When viewing each compute kernel as a ``black-box,'' as many
streaming systems do (e.g., RaftLib~\cite{bc15b,raftpage}), then 
buffer sizing is one of the most influential knobs available to tune the 
application aside from resource selection (i.e., the hardware the kernel
is executing on).
The sizing of each 
buffer (queue) within a streaming system has very
real performance implications (see Figure~\ref{fig:toobigalloc}).  Too small 
of a buffer will result in it always being full,  stifling performance 
of upstream compute nodes.  On the other hand, bigger buffers are 
not always better.  Extremely large buffers increase the overhead
associated with accessing them.
Discounting increased allocation time, excessively sized buffers 
can lead to increased page faults and virtual memory usage which 
can decrease performance.  Large buffers are also wasteful, 
physical memory is often at a premium,
it should be used wisely.  

In addition to being useful for queue (buffer) sizing, service rate observation is 
also critical for another optimization tasks.  Parallelization decisions
are often made statically, however streaming systems have the advantage
of state compartmentalization.  Compartmentalization simplifies parallelization
logic, methods such as that described by Gordon et al.~\cite{Gordon06}
or Li et al.~\cite{labc13} can
be used at runtime to increase parallelism and improve throughput. 
Platforms such as RaftLib have the capability of making parallelization decisions
both statically and at run-time.  One factor to consider when parallelizing
a compute kernel is the effective service rate of the kernel itself.  Knowledge
of the downstream compute kernels' service rates inform the run-time as to
if the system downstream is capable of accepting more input.  Knowing
the downstream kernel's non-blocking service rate is exactly what we need 
to know to make an informed parallelization decision.  To the best of our
knowledge there have been no other low overhead approaches to determine 
the online service rate of a compute kernel executing within a streaming
system.

In the sections that follow we will first describe some background 
material and related work, followed by a description of our heuristic
approach and it's implementation within the RaftLib streaming library.  
The heuristic's performance is evaluated with hundreds of micro-benchmark
executions, and two real world application examples implemented using
the RaftLib streaming framework.  Finally we will describe how this 
method can be combined with methods of online moment approximation to 
potentially estimate the shape of the underlying kernel's process 
distribution.

\section{Background \& Related Work}\label{sec:background}

At it's core, this work is about low-overhead instrumentation
of software systems.  The same techniques could also be 
used for co-designed hardware/software systems.  Many
others have produced low overhead instrumentation 
systems.  Amongst the earliest type of performance oriented instrumentation
tools were call graph
tools such as \texttt{gprof}~\cite{graham1982gprof}.
Other instrumentation tools such as TAU~\cite{shende2006tau} 
provide low overhead instrumentation and visualization for
MPI style systems. What these tools don't provide is a mechanism
for reporting performance during execution (which our system does).

Modern stream processing systems such as RaftLib~\cite{bc15b,raftpage}
can dynamically re-optimize in response to changing
conditions (workload and/or computing environment).  
To re-optimize buffer allocations there are generally two 
choices, either branch and bound search or analytic 
queueing model.  Branch and bound search has the 
disadvantage of requiring multiple allocations and re-allocations
until a semi-optimal buffer size is found.  Analytic queuing 
models are highly desirable for this purpose since
they can divine a buffer size directly, eschewing 
many unnecessary buffer re-allocations.  Compute kernel
mean service rates are, at a minimum, typically required
for these types of models.  Utilizing these
models dynamically therefore requires dynamic 
instrumentation.  Tools such as DTrace~\cite{cantrill2004dynamic},
Pin~\cite{luk2005pin}, and even analysis tools such
as Valgrind~\cite{nethercote2007valgrind} can provide
certain levels of dynamic information on executing
threads.  Our approaches differ from the aforementioned
ones in that we are specifically targeting
methods for estimating online service rate in a low overhead
manner.

Optimization of the queueing network inherent in stream
processing systems through optimal buffer sizing is only
one use of online service rate estimation, parallelization
control can benefit
as well. Parallelization 
decisions on streaming systems currently suffer 
from a lack of dynamic knowledge, that is, there
really isn't a way of knowing how parallelizing
a kernel will effect the overall performance of
an application.  When combined with methods 
described by Beard and Chamberlain~\cite{bc13b},
knowledge of online service rates can quickly
inform the run-time of how duplication will effect
the overall application's throughput.  

Work by Lancaster et al.~\cite{lwc11} laid out logic that could ostensibly
make online service rate determination possible.
They suggest measuring the throughput into a kernel when
there is sufficient data available within it's input queue(s) and
no back-pressure from its output queue(s).
This logic works well for FPGA-based systems where hardware is controlled
by the developer.  For multi-core systems, however, this logic
breaks down,  for several reasons which are enumerated
within this section.  The need 
for low overhead online service rate determination
motivates this work.

\begin{figure}[ht]
\centering
\includegraphics[height=2.5in,width=3.0in]{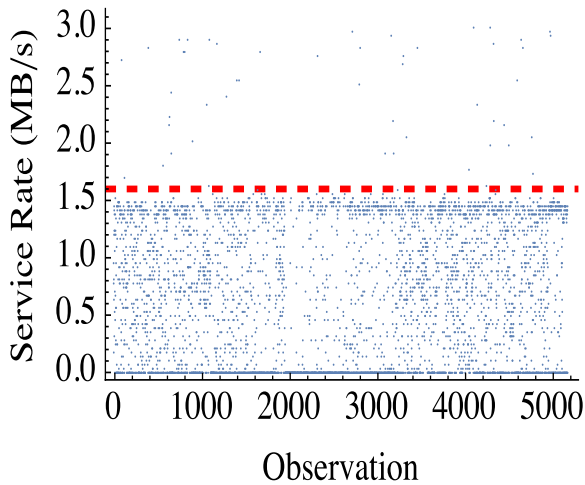}
\caption{Direct observations of the service
rate, using the logic of~\protect\cite{lwc11}, for a nominally
fixed rate microbenchmark kernel.
The x-axis is the increasing observation index with time, the y-axis represents
the actual data rate observed at each sample point.  The red dashed line
is the expected service rate as set experimentally.}
\label{fig:serviceratesample}
\end{figure}

The work of Lancaster et al.~assumes that the measurements
of a non-blocked service rate are all equal (i.e., the full service rate is 
observed at every sample point).  In reality things like partially
full queues result in less than realized service rates using this procedure. 
Further testing reveals that anomalies such as cache behavior and clock
variations can further exacerbate understanding of the true service
rate of a compute kernel.  Making things worse still are context
swaps that occur when one independent thread is observing another.
In reality, sampling the service rate of a compute kernel looks like 
Figure~\ref{fig:serviceratesample} where multiple outliers and noise 
confound our understanding of the true service rate.  

\begin{figure}[hbt]
\centering
\includegraphics[width=2.84in,height=1.81in]{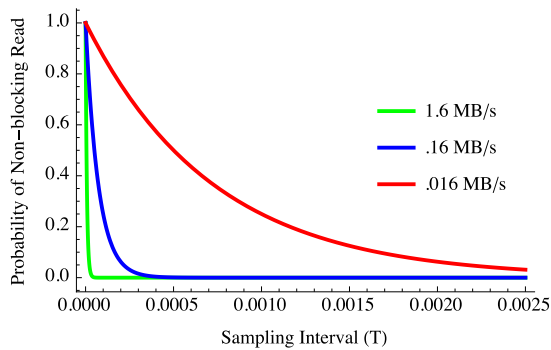}
\caption{The probability (y-axis) of observing a non-blocking read given
the observation period $T$ (x-axis).  In general the faster the server
or greater throughput the lower the probability of observing a non-blocking
read from the queue.}
\label{fig:prob}
\end{figure}

Central to accurate estimation of service rate is observing 
non-blocking reads and writes performed by the server.  While executing,
the probability of observing a non-blocked read or 
write to a queue in general is very low for high performance systems
(i.e., those systems whose compute kernels have high utilization).  
Equation~\ref{eq:prob} (a modification of the equations given by
Kleinrock~\cite{kleinrock1975queueing}) gives these probabilities for 
the simplified case where each server's process is Poisson
and only a single in-bound queue and out-bound queue are considered
(also known as an $M/M/1$ queue, using Kendall's
notation~\cite{kendall1957dictionary}).  Equation~\ref{eq:probread} 
is fairly intuitive: with $k = \lceil \mu_s T \rceil$, $k$ is
the mean number of items to be consumed by the server,
what is the probability of $n \ge k$ items existing in the in-bound queue 
given the service rate of the server ($\mu_s$) and the time period ($T$) over 
which the transactions are observed?  Just as intuitive is the estimate 
for the out-bound queue: for the server to have a non-blocking writes over
the entire time period $T$ then the queue must have space for that entire
time period, with $k = C - \lceil \mu_s T \rceil + 1$ and $k$ is the space
required by the server's output, the probability that $n < k$ 
(here, $n$ is the number of items in the out-bound queue) is 
given by Equation~\ref{eq:probwrite}.  Table~\ref{table:nomenclature} gives 
the list variable definitions.  Figure~\ref{fig:prob} shows this graphically for
a selection of throughput rates. In general the shorter the
service time, the lower the probability of observing a non-blocking
read or write.  Lengthening the observation period,
$T$, decreases the probability that blocking will not occur during 
the observation period whereas shorter periods increase the
probability of observation (i.e., no blocking during the period).

\begin{table}
\center
\caption{Nomenclature used for Equation~\protect\ref{eq:prob}}
\label{table:nomenclature}
\begin{tabular}{c|c}
Symbol & Description \\
\hline
$\mu_s$  & mean service rate \\
\hline
$\rho$ & server utilization \\
\hline
$C$    & capacity of output queue \\
\hline
$T$    & sampling period of monitor \\
\hline
$k$    & items needed by server during $T$ \\
\end{tabular}
\end{table}

\begin{subequations}
\label{eq:prob}
\begin{align}
k = \lceil \mu_s T \rceil \\
\label{eq:items}
Pr_{\mbox{\sc{\small{read}}}}(T,\rho,\mu_s) =  \rho^{k} \\
\label{eq:probread}\\
Pr_{\mbox{\sc{\small{write}}}}(T,C,\rho,\mu_s) = 
\begin{cases}
1 - \rho^{C - k + 1} & C \ge \mu_s T \\
0 & C < \mu_s T 
\end{cases}
\label{eq:probwrite}
\end{align}
\end{subequations}

The various mechanics required to estimate an online service rate will
be covered in the subsequent sections.  We will begin by describing the 
monitoring system itself at an abstract level, this is followed by setting
the observation period and finally the heuristic for online service rate
determination itself.

\section{Monitoring Mechanism} \label{sec:mechanism}

The simple act of observing a rate can change the behavior being observed.  
This phenomena is more obvious with large real world observations (e.g., 
observing animal behavior), however it is equally true for micro ones.  
Data observation might not make the data run away, however each observation 
requires non-zero perturbation to record it (e.g., a copy from the 
incremented register at some interval). Alternatively, saving every event
under observation can quickly overwhelm the hardware and operating system.
Trace files, even when compressed, can grow rapidly.  Determining the 
service rate with trace data in a streaming fashion (saving none of it)
might be possible, however it still increases traffic within the memory subsystem
which is less than desirable in high performance applications.  Concomitant 
to reducing communications overhead associated with monitoring is moving
any computation associated with that instrumentation out of the application's 
critical path.  To accomplish this, our instrumentation scheme (implemented 
within RaftLib) uses a separate monitoring thread.  This increases the 
sensitivity to timing precision and the possibility of noise within
the observations.  Within this section the overall architecture of our
implementation is discussed.

\begin{figure}[hbt]
\centering
\includegraphics[width=3in,height=3in]{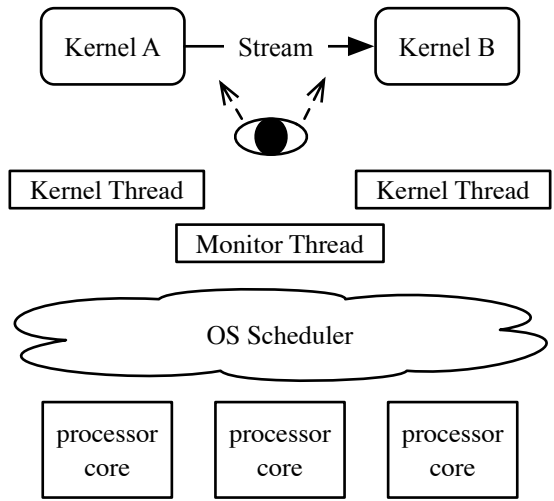}
\caption{High level depiction of the abstraction layers
coalesced around a simple streaming application with two
compute kernels.  An independent monitor thread serves
to instrument the queue.  Both the kernel threads and
monitor threads are subject to the runtime and operating
system (OS) scheduler.}
\label{fig:arrangement}
\end{figure}

At a high level, Figure~\ref{fig:arrangement} depicts
the arrangement of the instrumentation system under
consideration.  A simplified streaming application
with only two kernels is shown, connected by a single
stream.  Each kernel is depicted as executing on an 
independent thread.  A monitor (depicted as an eye), performs all
the instrumentation work, it executes on an independent
thread as well.  Each of these threads is scheduled
by the streaming run-time and the operating system.  Both
provide input on when each kernel and the monitor is
to execute.  Each of these threads also could execute on
independent processor cores or a single multiplexed core.
Each abstraction layer has the potential to impart noise
on any observations made by the monitor, the methods
proposed here must deal with and operate
in spite of this complexity.

To minimize overall impact, the data necessary to estimate
the service rate is split between the queue itself and 
the monitor thread.  This has the benefit of transmitting 
data only when absolutely necessary.  As depicted in Figure~\ref{fig:arrangement},
the queue itself is now visible to three distinct threads:
the monitor thread and the producer/consumer threads at 
either terminus of the queue.  The only logic to consider
within the queue itself is that necessary to tell the 
monitor thread if it has blocked and that necessary to increment
a item counter as items are read from or written to the queue.  The
monitor thread reads these variables written by the run-time
controlling the queue (work is actually performed by the 
producer or consumer threads).  The monitor thread also
resets or zeros the counter (which will be called $tc$ from
this point forward) and blocking boolean  kept by the queue.  In 
a non-locking operation, the monitor thread copies and
zeros $tc$.  This has the advantage of being
quite fast, however there are implications.

The monitor thread samples at a fixed interval of time $T$
which is the sampling period.  When the monitor thread
samples $tc$ and the blocking boolean, it has no way of 
knowing if the server at either end only performed complete
executions or partial
ones.  The only thing it can be certain of is that the data
read are non-blocking if the boolean value is set appropriately.  
This means that $tc$ can represent something less 
than the actual service rate.  Also
contained within the $tc$ are effects not-representative of average
behavior; these include (list not exhaustive): caching effects,
interrupts, memory contention, faults, etc.  

As mentioned previously (see Figure~\ref{fig:prob}), the
probability of making a non-blocking observation is in general
quite low.  In order to improve those odds there
are some mechanisms that the run-time can implement.  Given a full 
out-bound queue, resizing the queue provides a brief window over 
which to observe fully non-blocking behavior.  
Given an empty 
in-bound queue there are three implementable actions:
(1)~increasing the number
in-bound servers feeding more arrivals to the queue,
(2)~changing execution hardware 
of the up-stream server can have the
same effect as the aforementioned approach, 
(3)~adjusting the scheduling frequency before observing full
service rate in order to 
fill the in-bound queue.  Our implementation within RaftLib utilizes only
the first approach for out-bound queues, future implementations will 
utilize all of the above further improving service rate determination.



%

\section{Service Rate Monitoring} \label{sec:heuristic}

Online estimation of service rate requires four basic
steps: fixing a stable sampling period $T$, sampling
only the correct states (expounded upon below),
reducing and
de-noising the data, then estimating the non-blocking
service rate.
The queueing system has a finite
number of states which are useful in estimating
the non-blocking service rate.  The most obvious states
to ignore are those where the in-bound or out-bound
queue is blocked (see Lancaster et al.).  As mentioned
in Section~\ref{sec:mechanism}
there are also data unrepresentative of the non-blocking service rate.
Raw data such 
as Figure~\ref{fig:serviceratesample} are initially collected, filtering this
data through the process described below 
produces a final usable result.  We start by describing how we determine a stable sampling
period, $T$, followed by the heuristic to process the raw read data, $tc$.
Symbols used in this section are summarized in Table~\ref{table:heuristic}.

\begin{table}[ht]
\center
\caption{Nomenclature used for Section~\protect\ref{sec:heuristic}}
\label{table:heuristic}
\begin{tabular}{c|c}
Symbol & Description \\
\hline
$T$  & sampling period \\
\hline
$tc$ & sum of non-blocking reads during $T$ \\
\hline
$S$    & windowed set of items $tc$ \\
\hline
$S'$    & Gaussian filtered set of $S$ \\
\hline
$q$     & $95^{th}$ quantile of $S'$ \\
\hline
$\bar{q}$ & population averaged $q$ \\
\hline
$d$ & bytes per data item \\
\end{tabular}
\end{table}

\subsection{Sampling Period Determination}
Each queue within a streaming application has it's own monitor thread.  
As such, each $T$ is queue specific, since each
instrumented queue is in a slightly differing environment.  
An initial requirement is a stable time 
reference across all utilized cores. The timing method
described by Beard and Chamberlain~\cite{bc14a} is employed
(our specific implementation uses the \texttt{x86} 
\texttt{rdtsc} instruction, but any sufficiently high resolution time
reference could be used).  This method
provides a stable and monotonically increasing time reference 
whose latency on most systems is approximately $50 - 300$~ns
across the cores.
Despite a relatively stable time reference, two 
trends complicate matters.  First, as service time
decreases, the probability of observing a non-blocking queue 
transaction decreases as well (see Equation~\ref{eq:prob}).  
Second, noise from the system and timing mechanism dominate 
for very small values of $T$ making observations 
unusable~\cite{cadzow1970discrete}. 

Given a common understanding of time across cores, we can
now proceed to choose a sampling period, $T$.
Modern computing
systems (multi-core processors, multiple system services, general purpose
operating system, etc.) introduce
some level of noise into the measurements~\cite{bc14a,mazouz2010study}.
Given that we are measuring service rates, a longer
sample period helps
to smooth out these disturbances.
However, we wish to observe kernel executions that are unimpeaded
by their environment (no blocking due to upstream or downstream
effects).  This pushes us towards a shorter sampling period.

\begin{figure}[hbt]
\centering
\includegraphics[width=3in,height=2.16in]{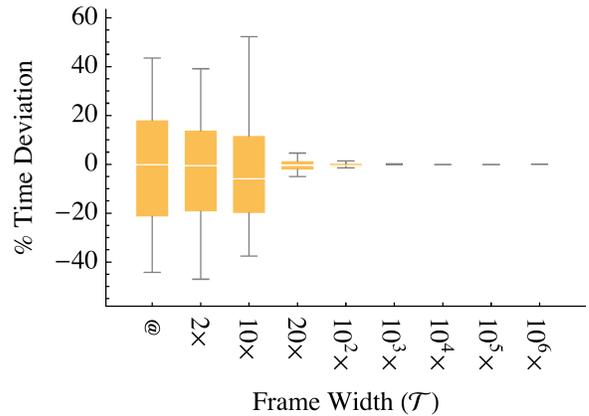}
\caption{Observations of $T$ variation using the timing
mechanism of~\cite{bc14a}.  The $@$ symbol represents
the minimum resolution of the timing mechanism ($\sim\! 300$~ns 
for this example), subsequent
box and whisker observations are the indicated multiple of $@$.
The trend indicates that wider time frames (up to the approximate
time quanta for the scheduler) give more stable values
of $T$.}
\label{fig:timing}
\end{figure}

Figure~\ref{fig:timing} shows how the measured (actual) sampling period
varies with desired sampling period, $T$, starting 
with the minimum latency ($\sim\! 300$~ns for this example) of back 
to back timing requests 
then iterating over multiples of that latency.  The monitor
thread tries to find the widest stable time period $T$ (moving to 
the right in Figure~\ref{fig:timing}) while minimizing
observed queue blockage during the period.  To make this more
concrete, our implementation lengthens the period if:
(1) no blockage occurred on the in-bound or out-bound buffer (with 
respect to a kernel) within the last $k$ periods and (2) the 
realized period of the monitor was within $\epsilon$ of the
current $T$ over the last $j$ periods (i.e., $T$ was stable).  
Failure to meet these conditions results in the failure of our method
(i.e., we conclude that our approach will not result in
usable service rate monitoring).

\subsection{Service Rate Heuristic}
Once a stable $T$ has been determined, the next step is estimating
the online service rate without having to store the entire
data trace of all queueing transactions. Each queue
stores transactional data at each end (referred
to as the head and tail).  The head and tail of the 
queue store counts of non-blocking transactions, $tc$, as well
as the size of each item copied, $d$.  The transaction count, $tc$,
requires very little overhead since it is simply
a counter.  The item size is typically constant for any
given queue.  The instrumentation thread samples $tc$ from
the head and tail of the queue every $T$ seconds.
While there are many factors that can slow down a kernel
in the context of a full execution, only a few can make
it appear to execute faster (see Figure~\ref{fig:serviceratesample}).
We will use an estimate of the maximum, well-behaved $tc$
to estimate the service rate of interest.
(What we mean by well-behaved is articulated below.)
For simplicity, the discussion that follows will consider only
actions that occur at the head of the queue (departures from
the queue into the server), with
the understanding that the same actions occur at the tail as well.

The overall process described below is summarized in
Algorithm~\ref{algo}.  This description
presumes
there is an implementaion of a streaming mean
and standard deviation (see Welford~\cite{welford1962note} and 
Chan et al.~\cite{chan1983algorithms}) 
through the $updateStats()$, $updateMeanQ()$
and $resetStats()$ methods.  Not described explicitly in the
algorithm is the convergence methodology described later in the text, which
is implemented within the $QConverged()$ function.  Padding is 
not used for the filter, therefore the filter starts at the 
radius for the filter so that the result of the filter has a 
width $2 \times \text{radius}$ smaller than the data window.  

\begin{algorithm}[ht]
 $stream \leftarrow tc$\;
 $output \leftarrow $ output stream\;
 $S \leftarrow \{\}$\;
 \While{True}{
   $tc_{current} \leftarrow pop( stream )$\;
   $S' \leftarrow \{\}$\;
   \For{ $i \leftarrow gauss_{radius}$, $i < |window| - gauss_{radius}$,$i++$ }{
      $val \leftarrow$ Dot( $S[ i - gauss_{radius} $;;$i + gauss_{radius} ]$,$GaussianFilter$ )\;
      push( $S'$, $val$ )\;
   }
   $\mu_{S'} \leftarrow  $Mean($S'$)\;
   $\sigma_{S'} \leftarrow $StandardDeviation( $S'$)\;
   $q \leftarrow $NQuantileFunction( $\mu_{S'}$, $\sigma_{S'}$, $.95$ )\;
   updateStats( $q$ );
   \If{ QConverged() }{
      push( $output$, getMeanQ() )\;
      resetStats()\;
   }
 }
 \caption{Service Rate Heuristic}
 \label{algo}
\end{algorithm}

While sampling $tc$, the timing thread creates an ordered list $S$,
where items are ordered by
entry time (easily implemented as a first-in first-out queue).
$S$ is maintained as a sliding window of size $w$.
If $S$ is of 
sufficient size, then it is expected
that the distribution of $tc \in S$ tends toward a Gaussian distribution
($\mathcal{N}(\mu_{S},\sigma_{S})$), as it is a list of sums of non-blocking
transactions.  $S$, however also consists of many data that 
are not necessarily indicative of the non-blocking service rates.
These elements arise from the following conditions: (1)~the monitor
thread observed only a partial firing of the server (i.e., the 
server had the capability to remove $j$ items from the queue
but only $<j$ items were evident when retrieving $tc$); (2)~the 
monitor thread clears the queue's current value of $tc$ during 
a firing (i.e., the counter maintaining $tc$ is non-locking because locking
it introduces delay); (3)~outlier conditions as
discussed in Section~\ref{sec:background} which are not indicative
of normal behavior conspire to speed up or slow down (momentarily)
the service rate.  We use a Gaussian filter to lessen the impact of these effects.

Filters are frequently used in signal processing applications to de-noise
data sets.  In general, a filter is a 
convolution between two distributions so that the response is
a combination of both functions.
The underlying
distribution of $S$ without outliers tends towards a Gaussian,
therefore a Gaussian discrete filter is used
to shape the data in $S$ so that it is sufficiently well-behaved (de-noised)
for estimating the maximum.  The filtered data make up the set $S'$.
The exact Gaussian kernel is described by Equation~\ref{eq:filter}, where
$x \gets [-2,2]$ is the index with respect to the center.  Through experimentation
a radius of two was selected as providing the best balance of 
fast computation and smoothing effect.  

\begin{equation}
GaussianFilterKernel(x) \gets \frac{e^{-\frac{x^2}{2}}}{\sqrt{2 \pi }}
\label{eq:filter}
\end{equation}

Once filtered, we use the data of $S'$ to estimate the maximum.
Since we must still account for outliers, rather than explicitly
use the sample maximum, we estimate the maximum of the well-behaved
counts via the $95^{th}$ quantile of $S'$.
This is a reasonable approximation given that: once filtered,
$S'$ even more closely has a Gaussian distribution than $S$
and a quantile is more robust to outliers than the sample maximum.
Operationally, we use the sample mean, $\widehat{\mu_{S'}}$, and
standard deviation, $\widehat{\sigma_{S'}}$, to estimate
$\mathcal{N}(\mu_{S'},\sigma_{S'})$, and the
quantile is
\begin{equation}
q = \widehat{\mu_{S'}} + 1.64485~\widehat{\sigma_{S'}}.
\label{eq:q}
\end{equation}

\begin{figure}[ht]
\centering
\includegraphics[height=2.5in,width=3in]{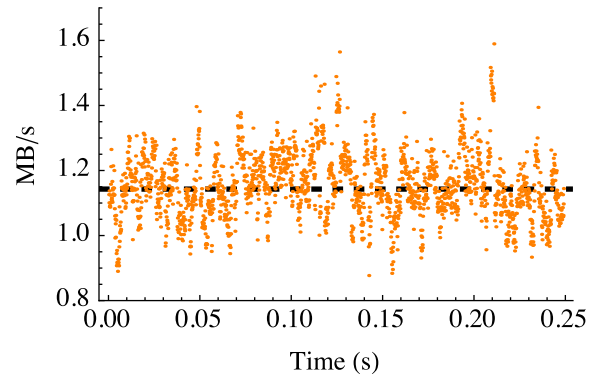}
\caption{Plot of the values of $q$ with increasing
time.  Each value of $q$ is the result of a 
computation of Equation~\protect\ref{eq:q}.  The dashed
line across the $y$-axis represents the set or expected
service rate.}
\label{fig:unstableq}
\end{figure}

\begin{figure}[ht]
\centering
\includegraphics[height=2.5in,width=3in]{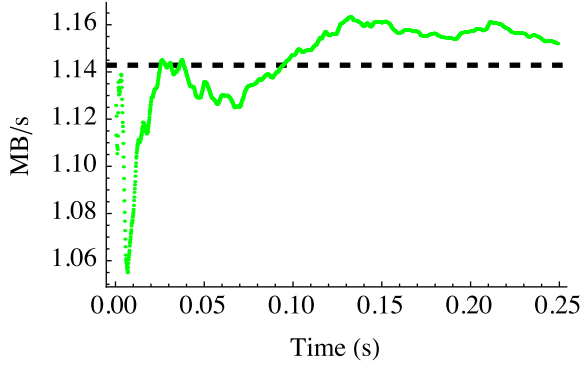}
\caption{An example of convergence of $\bar{q}$ with
increasing time.  This data is from a single queue tandem
server micro-benchmark, observing the departure rate from
the queue to the server with the set service rate marked
as a dashed line.}
\label{fig:stableq}
\end{figure}

Direct utilization of $q$ is sufficient for some purposes,
however it is only valid for the time period comprising the
window over which it was collected $p \gets T \times w$.  Subsequent
sets $S'_{i}$ update $\mu_{S'}$ and $\sigma_{S'}$ resulting
in frequent new values (e.g., Figure~\ref{fig:unstableq}).  
Stability is gained by using the online
mean of successive values of $q_{i}$.  Where $\bar{q}$ is 
the averaged, estimated maximum non-blocking transaction count $tc$,
assuming only one queue for simplicity, the service rate is simply 
$\frac{\bar{q} \times d}{T}$.  This, however also assumes
that the underlying distribution generating $tc$ is also 
stable.  As with all online estimates, that of $\bar{q}$ 
becomes more stable with more observations (e.g., Figure~\ref{fig:stableq}).

Convergence of~$\bar{q}$~to a ``stable'' value is expected after 
a sufficiently large number of observations.
In practice, with micro-second level sampling,
convergence is rarely an issue at steady state.
Determining when $\bar{q}$ is stable is accomplished
by observing $\sigma$ of
$\bar{q}$.  Minimizing the standard deviation is equivalent to 
minimizing the error of $\bar{q}$.  With a finite number of 
samples, it is unlikely that $\sigma(\bar{q})$ will ever
equal to zero, however observing the rate of change of the
error term to a given tolerance close to zero is a typical approach.
A similar windowed approach as that taken above
is used, however with differing filters to approximate the
relative change over the window.   A discrete 
Gaussian filter with a radius of one is followed by a Laplacian 
filter with discretized values (in practice, one combined filter
is used).  This type of filter is widely used in image edge 
detection.  Here, we are utilizing to minimize the standard 
deviation; essentially the filter gives a quantitative metric
for the rate of change of surrounding values.  The exact 
kernel is given in Equation~\ref{eq:laplacian} with $x \gets [-1,1]$
and $\sigma \gets \frac{1}{2}$.  The values of the minimum and maximum of the filtered 
$\sigma(\bar{q})$ are kept over a window $w \gets 16$
where convergence is judged by these values all being
within some tolerance (ours set to $5\times10^{-7}$).

\begin{equation}
LaplacianGaussian(x)\gets\frac{x^2 e^{-\frac{x^2}{2 \sigma ^2}}}{\sqrt{2 \pi } \sigma ^5}-\frac{e^{-\frac{x^2}{2 \sigma ^2}}}{\sqrt{2 \pi } \sigma ^3}
\label{eq:laplacian}
\end{equation}

\begin{figure}[ht]
\centering
\includegraphics[height=2.5in,width=3in]{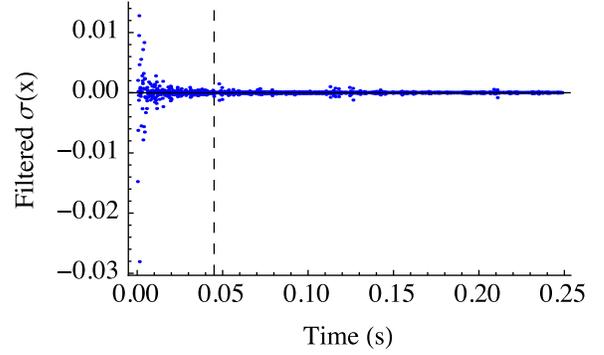}
\caption{Plot of the filtered standard deviation 
of $\bar{q}$, the point of convergence is indicated by the
vertical dashed line.}
\label{fig:convergedq}
\end{figure}

An example of a stable and converged $\bar{q}$ is shown
in Figure~\ref{fig:convergedq}, where the data plot is of
the dual filtered $\sigma(\bar{q})$ and the vertical line is the
point of convergence.  The time scale on the $x$-axis is
the same as that of Figure~\ref{fig:stableq} so that the stability
point on Figure~\ref{fig:convergedq} matches that of Figure~\ref{fig:stableq}.

\begin{figure}[ht]
\centering
\includegraphics[height=2.5in,width=3in]{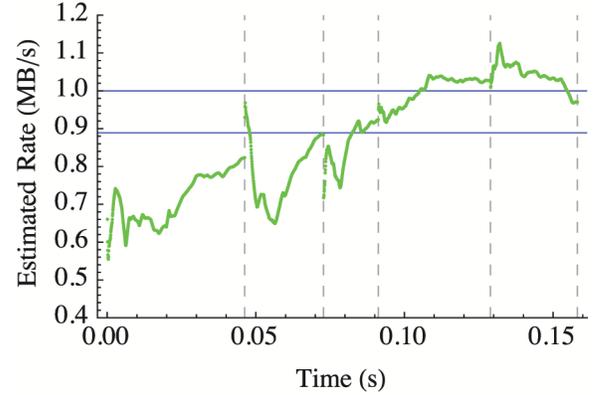}
\caption{Example of $\bar{q}$ adapting to two service
rates during execution of a micro-benchmark.  The instrumentation
captures the departure rate from a single queue to a compute 
kernel.}
\label{fig:dualphaseex}
\end{figure}

Once convergence is achieved, it is a simple matter to 
restart the process described above, finding a new value
for $\bar{q}$. Figure~\ref{fig:dualphaseex} shows 
a sample run where the departure rate of data elements from
a queue to the compute kernel.  Within this figure the actual
service rate is known (solid blue $y$-axis grid lines).  
The $x$-axis grid lines (dashed vertical lines) show where
our method has converged to a stable solution and re-started.
Changes in $\bar{q}$ are assumed
to mean a change in the process distribution 
governing $tc$.


\section{Experimental Setup}
\label{sec:methods}
\subsection{Infrastructure}
The hardware used for all empirical evaluation is listed
in Table~\ref{table:hardware}.  All code is compiled with the
``-O2'' compiler flag using the GNU GCC compiler (version 4.8.3).

In order to assess our method over a wide range of conditions,
a simple micro-benchmark consisting of two threads connected
by a lock-free queue is used.  Each thread consists of a while
loop that consumes a fixed amount of time in order to simulate
work with a known service rate.  The amount of work, or service-rate,
is generated using a random number generator sourced from
the GNU Scientific Library~\cite{galassi954161734gnu}.  The service
rates of kernels within the micro-benchmark are limited to approximately
$\sim 8$~MB/s due to the overhead of the random number generator and the 
size of the output item (8~bytes).  Service time distributions
are set as either exponential or deterministic.
Parameterization of the distributions
is selected using a pseudorandom number source.
The exact parameterization range and distribution are noted
where applicable within the results section.

\begin{table*}[htb]
\centering
\caption{Summary of hardware used for empirical evaluation}
\label{table:hardware}
\begin{tabular}{ c | c | c | c }
Platform & Processor & OS & Main Memory \\
\hline
1 & 2 $\times$ AMD Opteron 6136 & Linux 2.6.32 & 62 GB \\
\hline
2 & 2 $\times$ Intel E5-2650 & Linux 2.6.32 & 62 GB \\
\hline
3 & 2 $\times$ Intel Xeon X5472 & Darwin 13.4.0 & 32 GB \\
\hline 
4 & 2 $\times$ Six-Core AMD Opteron 2435 & Linux 3.10.37 & 32 GB \\
\hline 
5 & Intel Xeon CPU E3-1225 & Linux 3.13.9 & 8 GB
\end{tabular}
\end{table*}

The streaming framework used
is the RaftLib C++ template library~\cite{raftpage}.  The 
RaftLib framework previously included instrumentation for static service
rate determination (i.e., by running each compute kernel 
individually with an infinite data source and infinitely large
output queue).  The functionality of online service rate determination
as described by this work has also been incorporated into the platform.

\subsection{Applications}\label{sec:applications}

In addition to the micro-benchmarks described above,
two full streaming applications are also explored.
The first, matrix multiply,
is a synchronous data flow application that is expected to have
relatively stable service rates.  The second is a string search
application that has variable rates.  Ground truth 
service rates for each kernel are determined by executing each
kernel offline and measuring the rates individually using a 
large resident memory data source (constructed for each kernel)
and ignoring the write pointers so that it simulates an infinite
output buffer.

\subsubsection{Matrix Multiply}\label{sec:matrixmultiply}
\begin{figure}[ht]
\centering%
\includegraphics[width=3.14in,height=1.95in]{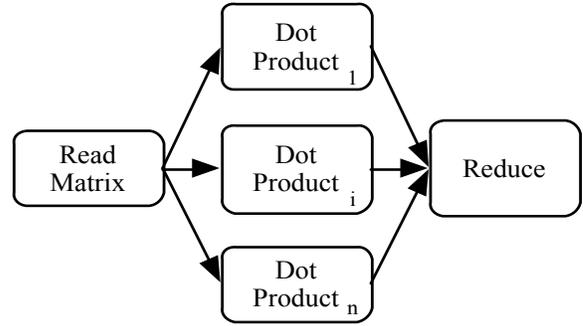}%
\caption{Matrix multiply application.  The first kernel reads both 
matrices to be multiplied and streams the data to an arbitrary ($n$) 
number of dot product kernels.  The final kernel reduces the 
input from the dot to a multiplied matrix.}%
\label{fig:matrixmultiply}%
\end{figure}

Matrix multiplication is central to many computing tasks. 
Implemented here is a simple dense matrix multiply ($C = AB$)
where the multiplication of matrices $A$ and $B$ are broken
into multiple dot-product operations.  The dot-product operation
is executed as a compute kernel with the matrix rows and columns
streamed to it.
This kernel can be duplicated $n$ times (see Figure~\ref{fig:matrixmultiply}).
The result is then streamed to a reducer kernel (at right) which 
re-forms the output matrix $C$.  This application differs
from the micro-benchmarks in that it uses real data read from
disk and performs multiple operations on it.  As with the 
micro-benchmarks, it has the advantage of having a continuous
output stream from both the matrix read and dot-product operations.

The data set used for the matrix multiply is a
$10,000 \times 10,000$ matrix of single precision floating
point numbers produced by a uniform random number generator.

\subsubsection{Rabin-Karp String Search}\label{sec:rabinkarp}
\begin{figure}[ht]
\centering%
\includegraphics[width=\columnwidth]{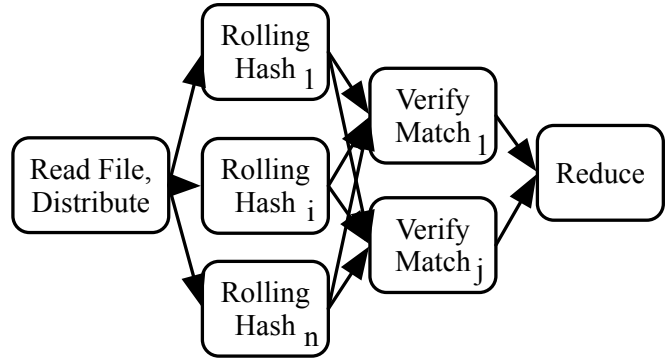}
\caption{Rabin-Karp matching algorithm.  The first compute 
kernel (at left) reads the file to be searched, hashes the
patterns to search and distributes the data to $n$ ``rolling-hash''
kernel(s).  Next are $j, j \le n$ verification kernel(s) to guard against 
matches due to hash 
collision. 
The final kernel (at right) is a reducer which consolidates 
all the results.}%
\label{fig:rabinkarp}%
\end{figure}

The Rabin-Karp~\cite{karp1987efficient} algorithm is classically used
to search a text for a set of patterns.  It utilizes a 
``rolling hash'' function to efficiently recompute the 
hash of the text being searched as it is streamed in.  The
implementation divides the text to be searched with an 
overlap of $m - 1$ (for a pattern length of $m$), so
that a match at the end of one pattern will not result
in a duplicate match on the next segment.  The output
of the rolling hash function is the byte position within
the text of the match.  The output of the rolling hash
kernel is variable (dependent on the number of matches),
for model selection testing purposes the input data will
be specially constructed in order to produce a regular 
steady state output.  
The next kernel verifies the match
from the rolling hash to ensure hash collisions don't 
cause spurious matches.  The verification matching kernel
can be duplicated up to $j$ times.  The final kernel
simply reduces the output from the verification kernel(s),
returning the byte position of each match (see
Figure~\ref{fig:rabinkarp}).  The corpus consists
of 2~GB of the string, ``foobar.''

\section{Results} \label{sec:results}

The methods that we have described are designed to enable
online service rate determination.  Just how well
do these methods work in real systems
while they are executing?  In order to evaluate
this quantitatively, several sets of micro-benchmarks
and real applications are instrumented to determine the
mean service rate of a given server.
We start with two sets of micro-benchmarks, the first
having a stationary distribution (with a fixed mean) and the second
having a bi-modal distribution that shifts its mean
halfway through its execution.

\begin{figure}[ht]
\centering
\includegraphics[height=2.5in,width=3in]{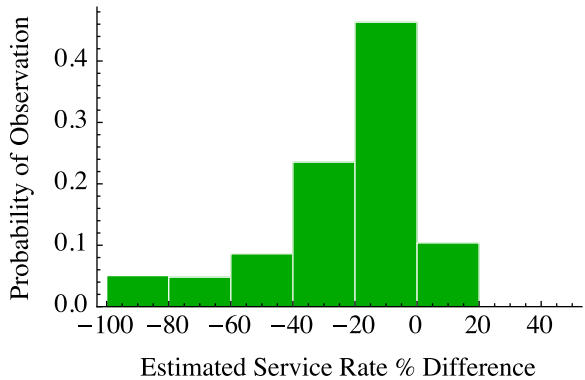}
\caption{Histogram of the probability of estimating the service
rate of Kernel B from Figure~\ref{fig:microbenchmark}.  Each execution 
is a data point, with the percent difference calculated as 
$(\frac{(\text{observed rate} - \text{set rate})}{\text{set rate}}) \times 100$.
Not plotted are four outliers to right of the plotted data which
are greater than $1000\%$ difference, which is not unexpected
given the probabilistic nature of our heuristic.}
\label{fig:microimageconverged}
\end{figure}

Each micro-benchmark is constructed with the configuration
depicted in Figure~\ref{fig:microbenchmark} and executed
with a fixed arrival process distribution.  The service
rate of Kernel B is varied for each execution of the
micro-benchmark from ($0.8$~MB/s~$\rightarrow\; \sim\! 8$~MB/s).
The results comprise $1800$ executions in total.
The departure rate from the queue is instrumented to 
observe the service rate of Kernel B.  The goal is 
to find the service rate of this kernel
without \textit{a priori} knowledge of the actual rate
(which we are setting for this controlled experiment). 
Figure~\ref{fig:microimageconverged} is a histogram of the
percent difference between the service rate estimated via
our method and the ``set'' filtered rate.

\begin{figure}[ht]
\centering
\includegraphics[height=2.5in,width=3in]{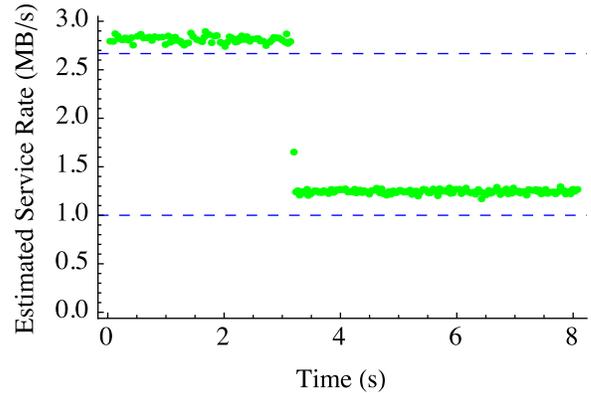}
\caption{Depiction of the ideal (drawn from empirical data) 
of the instrumentation's ability to estimate the service rate
while the application is executing.  Each dot represents the 
converged service rate estimate ($y$-axis).  The top and bottom
dashed lines represent the first and second phases as verified
by manual measurement in isolation.}
\label{fig:phasedist}
\end{figure}

We see in this histogram that generally the correspondence between
estimated service rate and ideal service rate is reasonably good.  We
expect divergence since these rates are determined while the application
is executing, not the full execution time average rate.  
When it errs, the estimate is typically low, which is consistent with
previous empirical data, in which actual realized execution times are
typically longer than nominal~\cite{bc14a}.
The majority of the results are within 20\% of nominal in any case.
The computational environment of any given kernel can 
change from moment to moment.  We simulate environment
change by moving the mean of the distribution halfway
through execution of Kernel B (with reference to the number of data elements
sent).  We are interested in whether our instrumentation can
detect this change, potentially enabling many online 
optimizations.  An ideal example with a wide switch in service rate is 
shown in Figure~\ref{fig:phasedist}
where the first phase is at $\sim\! 2.66$~MB/s
and the second is much lower at $\sim\! 1$~MB/s.  Not all examples are
so clear cut.

\begin{figure*}[t]
\centering
\includegraphics[width=6.57in,height=2.8in]{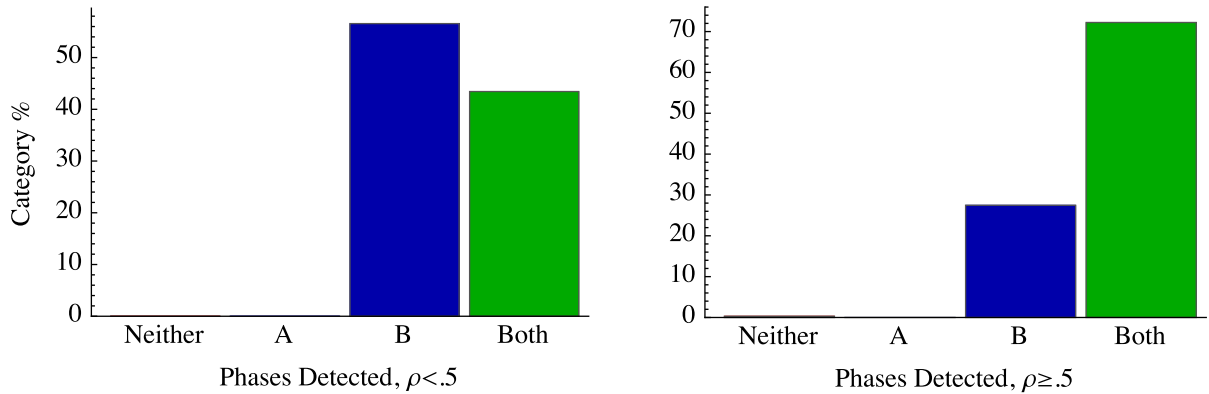}
\caption{Data from a dual-phase micro-benchmark that generates
two distinct service rate phases separated by server utilization,
$\rho$, and then by correct classification at each phase 
(as ``Neither,'' ``A,'' ``B,'' or ``Both''), which 
represent the heuristic finding none, only the first phase,
only the second phase or both, respectively.}
\label{fig:dualphaserho}
\end{figure*}

In order to classify the dual phase 
results into categories, a percent difference ($20\%$) from the manually
determined rates for each phase is used. Approximately 
$14.7\%$ of the data had nominal service rate shifts that 
were known to be less than the $20\%$ criteria specified.
Figure~\ref{fig:dualphaserho} shows the effectiveness of
our technique in categorizing the distinct execution phases
of the micro-benchmarks.
The rightmost graph shows the categorizations for low $\rho$,
and the leftmost graph shows the categorizations for high $\rho$.
Here, we make two observations.  First, the system correctly
detects both phases more effectively in high utilization conditions,
which are the conditions under which correct classification
is likely to be more important.  Second, the classification
errors that are made are all conservative.  That is, it is
correctly detecting the final condition of the kernel, indicative
of a conservative settling period for rate estimation.

It is well understood that a server with sufficient data on it's input
queue should be able to proceed with processing (assuming
no other complicating factors).  Therefore one trend that
we expect to see is an improvement in the approximation
for higher server utilizations.  In addition, servers that are 
more highly utilized typically have a much more profound impact 
on the performance of the application as a whole (e.g., they 
are dramatically more likely
to be throughput bottlenecks in the overall data flow).


Overall, the heuristic did quite well.  Looking at the single
phase data, only four of the micro-benchmark results were
extremely off. The dual phase data were also fairly good,
the heuristic failed to find either phase in only $0.24\%$ of the instances.
The real test of any instrumentation is how well it
can handle situations beyond those that are carefully controlled.
The only variable that is within the users' control is that
of data set selection.  Notably these applications are
not limited to the slower service rates of the micro-benchmark 
applications but are dependent on the mechanics of the application.  
The matrix multiply application is executed
on platform 2 from Table~\ref{table:hardware} with the number
of parallel dot-products set to five.  Only the reduce kernel 
is instrumented (see Figure~\ref{fig:matrixmultiply}) as 
the dot-products would be rather easy given the high data rates
inherent in transmitting an entire row by copy.  The ground-truth service
rate realized by each queue (the total service rate being a 
combination of rates from each input-queue) are determined by 
the method described in Section~\ref{sec:matrixmultiply}.
Overall the results are not quite as clean as those of the 
microbenchmark, but that is expected given the chosen kernel 
has an extremely low $\rho$.  A majority ($63\%$)
are within the range of measurements observed during manual 
estimation removing each kernel
from the system and manually measuring data rates at each input
port).

\begin{figure}[ht]
\centering
\includegraphics[height=2.5in,width=3in]{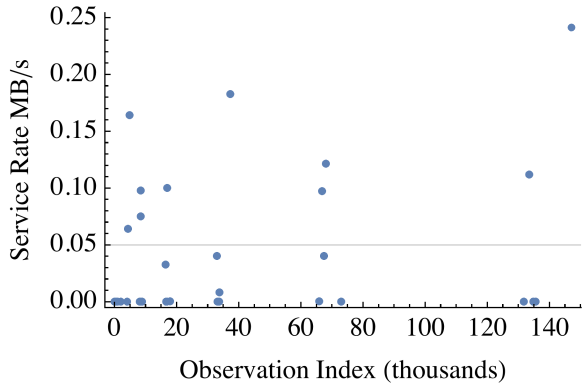}
\caption{Plot of the trace for the instrumented partial
service rate of the reduce kernel (the full rate being
the sum of all rates for each in-bound queue).  The 
manually determined rate for this experimental setup ranged
from $0.05$~MB/s to $0.43$~MB/s.  Overall, a majority of the results
$\sim\! 63\%$ are within this range.}
\label{fig:matrixmultiplyresults}
\end{figure}

\begin{figure}[ht]
\centering
\includegraphics[height=2.5in,width=3in]{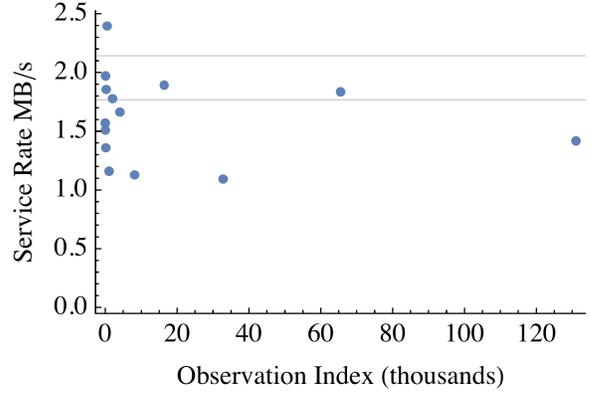}
\caption{Plot of the converged estimates of service rate for a 
single queue within the Rabin-Karp string matching application.
The utilization of
this server is less than $0.1$ meaning that the queue is almost 
always empty which leads to less opportunity for recording non-blocking
reads from the queue.}
\label{fig:rkhist}
\end{figure}

Similar to the results for the matrix-multiply application, 
the results for the Rabin-Karp application are also 
relatively good (recall that these are rates taken at points over
the course of execution).  The application is executed on platform 2 
from Table~\ref{table:hardware} with the number of matching
kernels fixed at four and the number of verification kernels
fixed to two.  Figure~\ref{fig:rkhist} shows the
online service rate by convergence point each data point represents a 
converged estimate of the service rate (potentially multiple
convergences for a single application execution).  Instrumented is a single
queue arriving to the verify block from the hash kernel.  Again, we've intentionally
picked a case where the $\rho$ is very low, which is very 
difficult for the instrumentation to find a non-blocking read
from the queue. In total, only $\sim\! 35\%$ of estimates are within 
the range observed when manually measuring service rate, although most 
of the data points are fairly close.  This highlights the limitations
of our approach.  If the non-blocking reads are not observed then
the rate simply cannot be determined with too much accuracy.

Low overhead instrumentation should be exactly that, low overhead.  This
means that there should be little, if any, impact on the execution of 
the application itself.  Low impact also means that the system executing both 
the application and instrumentation should see as little increase in 
overhead as possible.  Given that our system utilizes a separate
monitor thread, this could be a concern.  Using the single queue
micro-benchmark, the impact was measured with instrumentation and 
without instrumentation.  Using the GNU \texttt{time} command over 
dozens of executions, the average impact is only $1$~-~$2\%$.  Impact
to the system overall was equally minimal, load average increased
only a small amount (by $0.1$ on average).

\section{Conclusions \& Future Work}\label{sec:conclude}

This paper proposes and demonstrates a heuristic that enables online 
service rate approximation of each compute kernel within
a streaming system.  In streaming systems that exhibit 
filtering, the heuristic presented here can also be used
to detect non-blocking departure rates which can inform
a run-time of routing decisions made by the kernel as well
as the amount of filtering currently exhibited.  Overall 
our methodology works quite well.
When the heuristic fails,
it usually fails knowingly (e.g., no convergence is reached or non-blocking
reads were not observed).  A few cases of non-convergence
are included in the results (and noted).
Currently the default in RaftLib
is to fall back on the current best solution, but note the 
non-converged state.
Future work might include the 
solutions mentioned in Section~\ref{sec:mechanism} (i.e., 
temporary throttling and duplication) which 
could improve the probability of convergence.

Alluded to earlier was the potential to estimate the 
likelihood of the distribution process modulating the data 
stream as matching a known distribution; which is quite
useful if the known distribution enables the use of a 
closed form modeling solution.  We've shown a heuristic 
that can estimate the central moment of the service
process and it's second moment (variance) in a streaming
manner (for these calculations, only saving sums and discarding
the actual values).  Efficient methods also exist for streaming
computation of higher moments~\cite{pebay2008formulas}.  
Using the method of moments along with some simple classification,
it should be clear that online distribution selection can be performed
using the techniques described within this work as a 
basis, then extending them to include higher moment estimation.
Future work, and extensions to the RaftLib instrumentation
system, will include these pieces.

Parallelization decisions can easily benefit
from the information that this method provides.
Instead of relying on static (compile time) information,
decisions can now be made with up-to-date 
data improving optimality of the execution.  Related,
but not shown here, is the ability of this 
process to instrument streams entering or 
exiting a TCP stack.  It is assumed that
there should be no difference in monitoring 
user-space queues feeding data into a TCP 
link.  An open question is exactly how best
to synchronize the ingress and egress transaction
data.

In conclusion, we've demonstrated a probabilistic heuristic
that under most conditions can estimate the service
rate of compute kernels executing within a streaming 
system while that application is executing.  It has
been demonstrated to be effective using micro-benchmarks
and two full stream processing applications on multi-core
processors.


\section*{Acknowledgments}
This work was supported by Exegy, Inc., and VelociData, Inc.
Washington University in St. Louis and R.~Chamberlain receive income
based on a license of technology by the university to Exegy, Inc.,
and VelociData, Inc.



\bibliographystyle{IEEEtranS}
\bibliography{paper}
%
%
%

\end{document}